\documentclass[journal]{IEEEtran}

\usepackage{xcolor,soul,framed}
\usepackage[cmex10]{amsmath}
\usepackage{array}
\usepackage{mdwmath}
\usepackage{amsmath}
\usepackage{mdwtab}
\usepackage{eqparbox}
\usepackage{url}
\usepackage{cancel}
\usepackage{siunitx}
\usepackage{subcaption}
\usepackage{dblfloatfix}
\usepackage[pdftex]{graphicx}
\makeatletter
\AtBeginDocument{\let\hl\@firstofone}
\makeatother

\begin{document}
\bstctlcite{IEEEexample:BSTcontrol}
    \title{Toward Memristor-like Resonant Sensors: Observation of Pinched Hysteresis within MEMS Resonators}
  \author{Erion~Uka,~\IEEEmembership{Graduate Student Member,~IEEE,}
      Chun~Zhao,~\IEEEmembership{Senior Member,~IEEE}%

  \thanks{Manuscript received June 2, 2025. This work was funded in part by the the Royal Society Grant RG\textbackslash R2\textbackslash 232112, in part by the EPSRC IAA Grant IFPT\textbackslash 2023\textbackslash 22, and in part by the School of Physics, Engineering and Technology, University of York PhD studentship.}
  \thanks{E. Uka and C. Zhao are with the School of Physics, Engineering, and Technology, University of York, Heslington, York, UK, YO10 5DD.}%
  \thanks{C. Zhao is also affiliated with the York Biomedical Research Institute, University of York, Heslington, York, UK, YO10 5DD.}%
  \thanks{Corresponding email: \emph{chun.zhao@york.ac.uk}.}
}  

\maketitle

\begin{abstract}
Memristors, uniquely characterized by their pinched hysteresis loop fingerprints, have attracted significant research interest over the past decade, due to their enormous potential for novel computation and artificial intelligence applications. Memristors are widely regarded as the fourth fundamental electrical component, with voltage and current being their input and output signals. In broader terms, similar pinched hysteresis behavior should also exist in other physical systems across domains (e.g., physical input and electrical output), hence linking the real physical world with the digital domain (e.g., in the form of a physical sensor). In this work, we report the first observation of pinched hysteresis behavior in a micro-electro-mechanical systems (MEMS) resonator device, showing that it is viable to create resonant MEMS sensors incorporating memristor-like properties, i.e., \textit{MemReSensor}. We envisage that this will lay the foundations for a new way of fusing MEMS with artificial intelligence (AI), such as creating in-physical-sensor computing, as well as in-sensor AI, e.g., multi-mode in-sensor matrix multiplication across domains. 
\end{abstract}
 

\IEEEpeerreviewmaketitle

\section{Introduction}
\IEEEPARstart{M}{emristors}, since their theoretical prediction in 1971 \cite{chua1971memristor} and demonstration in 2008 \cite{strukov2008missing}, are emerging as a key building block for next-generation high-efficiency brain-inspired computing \cite{zhang2020brain} and AI hardware \cite{huang2024memristor}. Regardless of scale, geometry, material choice or design, memristors are defined by a key property, known as the ``\emph{pinched hysteresis}" \cite{chua2014if}. In a broader sense, memcapacitors and meminductors are also defined based on the pinched hysteresis property \cite{yin2015memristor}, as well as memtransistors \cite{sangwan2018multi}, which have all shown great promise in related computation and AI hardware applications \cite{demasius2021energy, wang2023beyond, yan2022progress}. It is worth pointing out that all of the aforementioned examples exhibit pinched hysteresis purely in the electrical domain, i.e., electrical input and electrical output.

Applying the same logic, and expanding on that mentioned above, it is intriguing whether this pinched hysteresis behavior can be observed across domains, e.g., physical input and electrical output. While this idea has been touched upon \cite{chiolerio2015ultraviolet, vahl2019concept}, the proposed devices, termed as ``memsensors", are essentially memristor devices, the pinched hysteresis behavior of which can be tuned by a physical input, i.e., incident light intensity. It has yet to be observed whether pinched hysteresis loops can be formed \emph{directly} between a physical input and the corresponding electrical output of a sensor. 

To realize this, a natural candidate is MEMS sensors, as they have been used extensively to transduce physical inputs into electrical outputs with high sensitivity and resolution \cite{wang2022resonant, pandit2019high}, boasting high reliability and low cost per device. Resonant MEMS have also been demonstrated as useful candidates for other interesting areas of research such as non-reciprocal devices e.g., isolators and circulators \cite{torunbalci2018fbar, yu2019highly, shao2020non}. Despite this, to date, we have not yet seen any demonstrations of MEMS-based sensors with the pinched hysteresis behavior.

In this work, we take the first step toward cross-domain memristor-like behavior, and will demonstrate the first pinched hysteresis loops observed within a generic silicon-based MEMS resonator device. This behavior and observation are enabled by applying \emph{parametric modulation signals} (PMSs) to dynamically couple multiple intrinsic vibration modes \cite{zhou2019dynamic, zhao2019toward}. Due to the unique phase properties of the MEMS resonator with coupled modes, when using a phase-locked loop (PLL) to lock onto the phase corresponding to resonant frequencies (in the exact same manner as in practical resonant sensors \cite{mustafazade2020vibrating}), the resonator can switch between multiple modes with hysteresis. When a single PMS is applied, two modes are coupled, and we can observe a pinched hysteresis loop between a stiffness perturbation and the output amplitude. When two PMSs are applied, three modes are virtually coupled, and a pinched hysteresis loop can be observed between a stiffness perturbation and the output frequency. We further extend our approach to apply three PMSs, and we can observe two pinched intermediate branches in the output frequency and three pinched hysteresis loops in the output amplitude, i.e., \emph{multiple pinched hysteresis} (MPH) in both the frequency and amplitude responses. Although the input in this work is still a stiffness perturbation due to an electrical voltage change, it has been demonstrated in the literature \cite{thiruvenkatanathan2009enhancing, de2016review}, that the electrical input essentially acts as a proxy for other stiffness-related physical inputs, suggesting that our approach can be easily translated to MEMS sensors with physical inputs.

As far as we are aware, this is the first demonstration of pinched hysteresis loops within MEMS devices, and so we have termed it the \emph{MemReSensor} (\underline{mem}ristor-like \underline{re}sonant \underline{sensor}). We believe that this discovery will pave the way for future generations of in-sensor computing (i.e., edge computing) \cite{zhou2020near}, neuromorphic computing (e.g., reservoir computing using hysteresis \cite{caremel2024hysteretic}), and even direct matrix multiplication from multi-physical-input vectors to electrical outputs.

\section{Theory}
\subsection{Parametric Modulation}

\begin{figure}[h!]
\centering
    \includegraphics[width=0.45\textwidth]{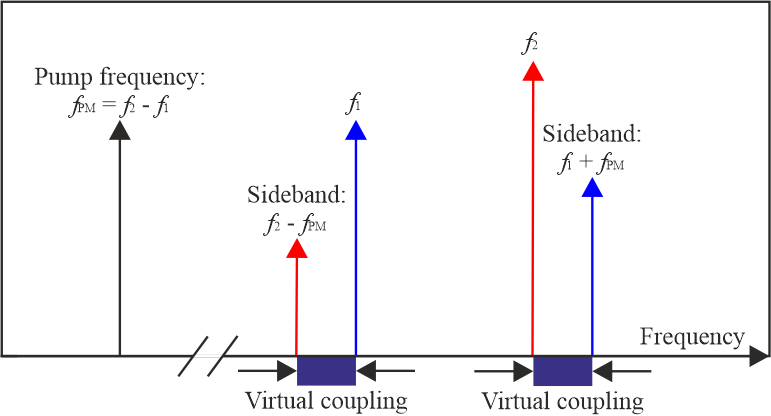}
    \caption{Illustration of virtual coupling when using one PMS. The intra-modal interactions when using a different number of PMSs can be visualized in the same way, where virtual coupling generates complex dynamics.}
    \label{PM_explanation}
\end{figure}

Parametric modulation is an operating scheme that generates periodic stiffness modulation in resonant MEMS devices using dynamic PMSs. As opposed to the recently proposed blue-sideband excitation \cite{xu2022closed, xi2023multiple, uka2024pt1}, here we focus on \emph{red-detuned} parametric modulation \cite{zhou2019dynamic, zhao2019toward} - where the pump frequency is close to the \emph{difference} between two modes of interest (see Eq. 2). This generates virtual coupling between the two modes \cite{zhang2020amplitude} (see Fig. \ref{PM_explanation}). The equations of motion describing the dynamics of an electrostatically transduced clamped-clamped (C-C) type MEMS resonator have been derived previously \cite{li2021enhancing}. These equations can also be used to describe the behavior of the electrostatically transduced DETF-type resonator used in this work. The generalized form of the equations of motion, which are suitable to describe dynamical systems under multiple PMSs, can be written as:

\begin{align}
\begin{split}
& \underbrace{m_{i}\ddot{u_{i}}(t) + c_{i}\dot{u_{i}}(t) + (k_{i}+k_{e, i})u_{i}(t)  
= F_{i}\cos(2\pi f_{\text{drive}}t)}_{\text{Linear equation of motion of the $i$th mode}} \\ 
&+ \underbrace{\sum_{\substack{j=1 \\ j\neq i}}^{n}\Bigg[k_{c}^{i,j}u_{j}(t)\Bigg]}_{\text{Linear coupling terms}} \\
&+\underbrace{\sum_{\substack{j=1 \\ j\neq i}}^{n} \Bigg[\cos(2\pi f_{p}^{i,j}t)\Bigg(\lambda_{p}^{i,j}u_{i}(t)+\Gamma_{p}^{i,j}u_{j}(t)\Bigg)\Bigg]}_{\text{Coupling terms due to parametric modulation}}
\end{split}
\label{eq:pmequations}
\end{align}

where $m_{i}$, $c_{i}$, $k_{i}$, $k_{e, i}$, $u_{i}(t)$, $F_{i}$ and $f_{\text{drive}}$ are the effective mass, damping, mechanical stiffness, electrostatic stiffness, deflection in time, driving force amplitude, and frequency of the drive signal, respectively. The number of modes ($n$) relates to the number of PMSs ($n-1$), e.g., one PMS couples two modes. $k_{c}^{i,j}$ is the linear coupling term between the $i$th and $j$th modes. $f_{p}^{i, j}$, $\lambda_{p}^{i, j}$ and $\Gamma_{p}^{i, j}$ are the frequency of the PMS used to couple the $i$th and $j$th modes, and the intra-modal and inter-modal pumping coefficient due to the resulting interactions between the $i$th and $j$th modes, respectively \cite{zhou2019dynamic}. 

An offset voltage ($v_{\text{offset}}$) is used here to introduce a stiffness perturbation in $k_{e,i}$ (see Eq. \ref{eq:pmequations}). In this work, this acts as a proxy for the stiffness perturbation caused by external physical inputs in resonant sensors, e.g., acceleration in resonant MEMS accelerometers \cite{mustafazade2020vibrating} or magnetic field strength in resonant MEMS magnetometers \cite{wang2022resonant}. 

Following \cite{li2021enhancing}, the theoretical equivalent stiffness change to the first mode ($f_{1}$) per volt can be calculated using the following equation: $2\phi_{1}V_{bias}\epsilon_{0}tl/d^{3}$, where $\phi_{1}$ is a coefficient required to modify the electrostatic stiffness for parallel plate, since the gap between the moving beam and the electrode is mode shape dependent. Substituting in $V_{bias} = 30$ V, $t = 25$ $\mu$m, $l = 384$ $\mu$m (the length of the electrode), $d = 2$ $\mu$m, and $\phi_{1}=0.197$ for the first mode \cite{li2021enhancing} gives $\Delta k_{e, 1}$/$\Delta v_{\text{offset}}$ $\approx0.126$ N/m/V. This agrees well with the estimation based on experimental data, which is $0.121$ N/m/V. This is calculated based on $\Delta k / 2k_{1} \approx \Delta f / f_{1}$ \cite{thiruvenkatanathan2009enhancing}, then substituting in values for $k_{1}\approx 99$ N/m obtained from COMSOL simulation, and $\Delta f\approx 63$ Hz (for 1 V of V\textsubscript{offset} change) and $f_{1}\approx 103.352$ kHz obtained from experiments.

\subsection{Simulated behavior of a one PMS system}
\begin{figure}[h!]
\centering
\begin{subfigure}{0.45\textwidth}
    \centering
    \hspace*{-1cm}\includegraphics[width=0.8\textwidth]{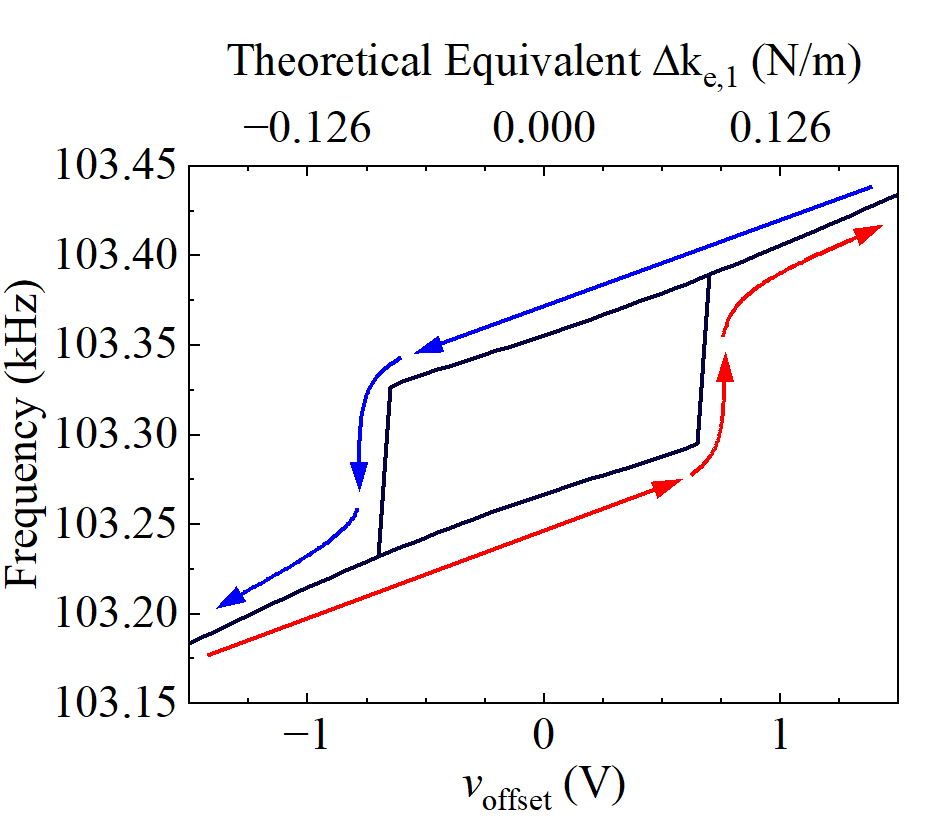}
    \caption{Simulated stiffness perturbation - frequency space}
    \label{sim_1PM_freq_resp}
\end{subfigure}%
\vspace{0.1 cm}
\begin{subfigure}{0.45\textwidth}
    \centering
    \includegraphics[width=0.8\textwidth]{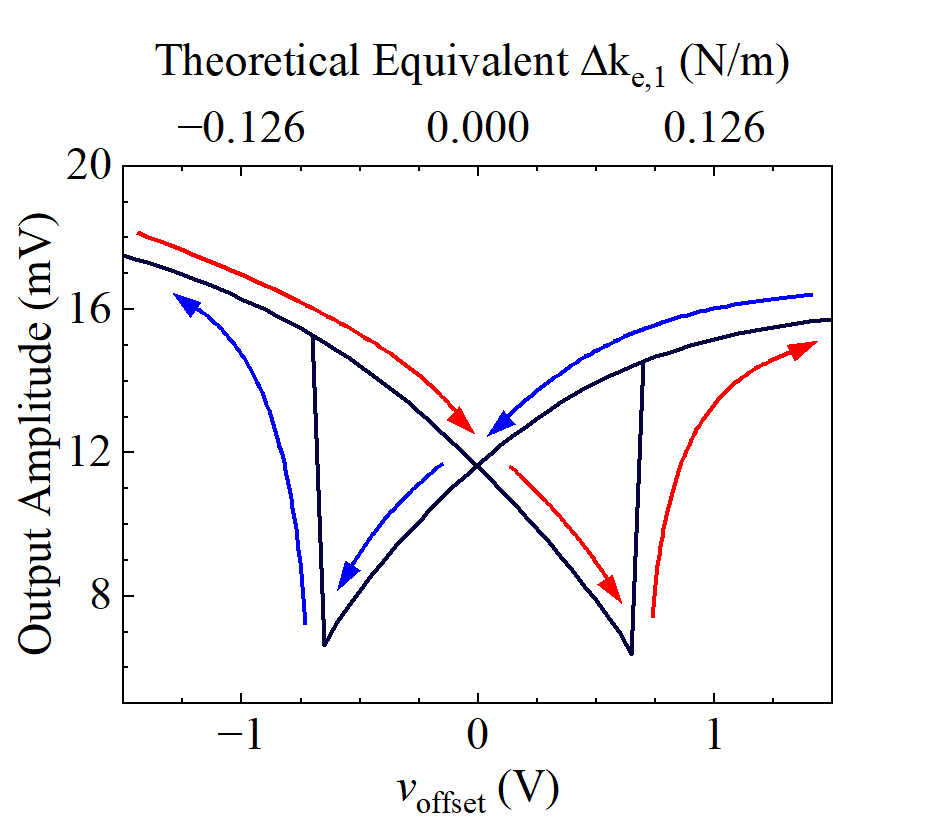}
    \caption{Simulated stiffness perturbation - amplitude space}
    \label{sim_1PM_amp_resp}
\end{subfigure}
% \vspace{0.1 cm}
\caption{The simulated (a) frequency - single hysteresis loop, and (b) amplitude - \emph{pinched hysteresis loop}, when exploring the virtual coupling between $f_{1}$ and $f_{3}$ with 1 PMS applied, using the values for the device as listed in Tab. \ref{tab:simulationparameters}}
\end{figure}

Considering the complexity of modeling the behavior of a slotted DETF resonator, we have extracted the following parameters based on a combination of experimental data and numerical estimations (see Tab. \ref{tab:simulationparameters}), which are considered a good representation of the parameters in the experiment. Using these parameters and Eq. \ref{eq:pmequations}, the dynamic behavior of the system, concerning the first and the third mode, with 1 PMS signal ($\omega_{p}^{1,3} = 2 \pi f_{p}^{1,3}$) applied, has been numerically simulated using the Harmonic Balance Method (HBM) \cite{kacem2011computational}. The simulated results are shown in Fig. \ref{sim_1PM_freq_resp} and \ref{sim_1PM_amp_resp}. It is worth pointing out that the computational demand scales rapidly with increased system complexity (e.g., increased number of PMSs and coupled modes or increased $n$), therefore only the 1 PMS case has been simulated in this work.

\renewcommand{\arraystretch}{1.5}
\begin{table}[h!]
    \caption{The values used to simulate the 1 PMS case}
    \centering
    \begin{tabular}{ c | c | c | c} 
     \hline
     Parameter & Value & Parameter & Value \\ 
     \hline 
     $m_{1}$ & \num{2.17 e-10} & $\lambda_{p}^{1, 3}$ & \num{0.38}\\ 
     
     $m_{3}$ & \num{9.06 e-10} & $\lambda_{p}^{3, 1}$ & \num{0.42} \\ 
     
     $c_{1}$ & \num{9.16 e-08} & $\Gamma_{p}^{1, 3}$ & \num{-0.33} \\ 
     
     $c_{3}$ & \num{2.12 e-07} & $\Gamma_{p}^{3, 1}$ & \num{-0.33} \\ 
     
     $k_{1}$ & \num{99.01} & $F_{1}$ & \num{2.82e-10} \\ 
     
     $k_{3}$ & \num{4983.70} & $F_{3}$ & \num{3.09e-10} \\  %$\eta_{1}$ & \num{1.66 e-07} \\ [1ex] 
     
     $k_{c}^{1, 3}$ & \num{0.22} & $f_{p}$ & \num{2.67 e+05} \\   %$\eta_{2}$ & \num{1.82 e-07} \\ [1ex] 
     
     $k_{e, 1}$ & \num{7.53} & $Q_{1}$ & \num{1.60 e3} \\ 
     
     $k_{e, 3}$ & \num{8.39} & $Q_{2}$ & \num{1.00 e4} \\  
     
     \hline
    \end{tabular}
    \label{tab:simulationparameters}
    \end{table}

\section{Experimental Setup}
\subsection{Device Design}
\begin{figure}[h!]
\centering
\begin{subfigure}{0.45\textwidth}
    \centering
    \includegraphics[width=0.8\textwidth]{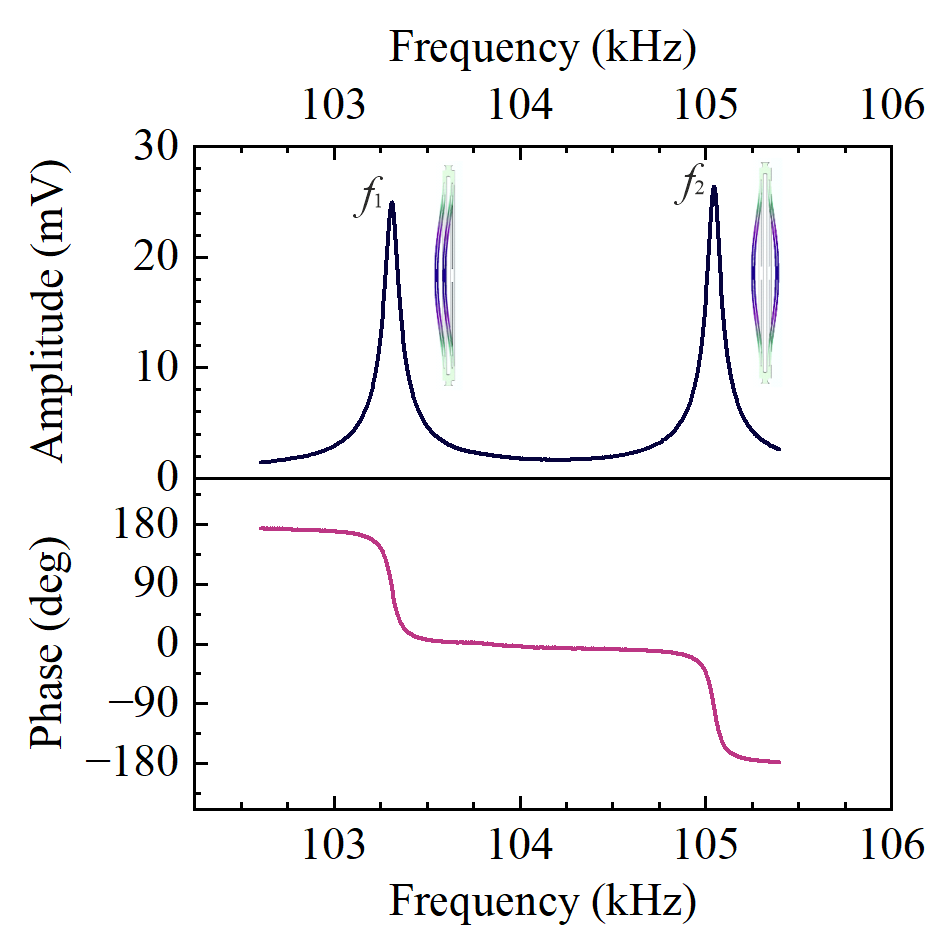}
    \caption{Frequency sweeps for modes $f_{1}$ and $f_{2}$.}
\label{linear_sweeps_f1_f2}
\end{subfigure}

\vspace{0.2cm}

\begin{subfigure}{0.45\textwidth}
    \centering
    \includegraphics[width=0.8\textwidth]{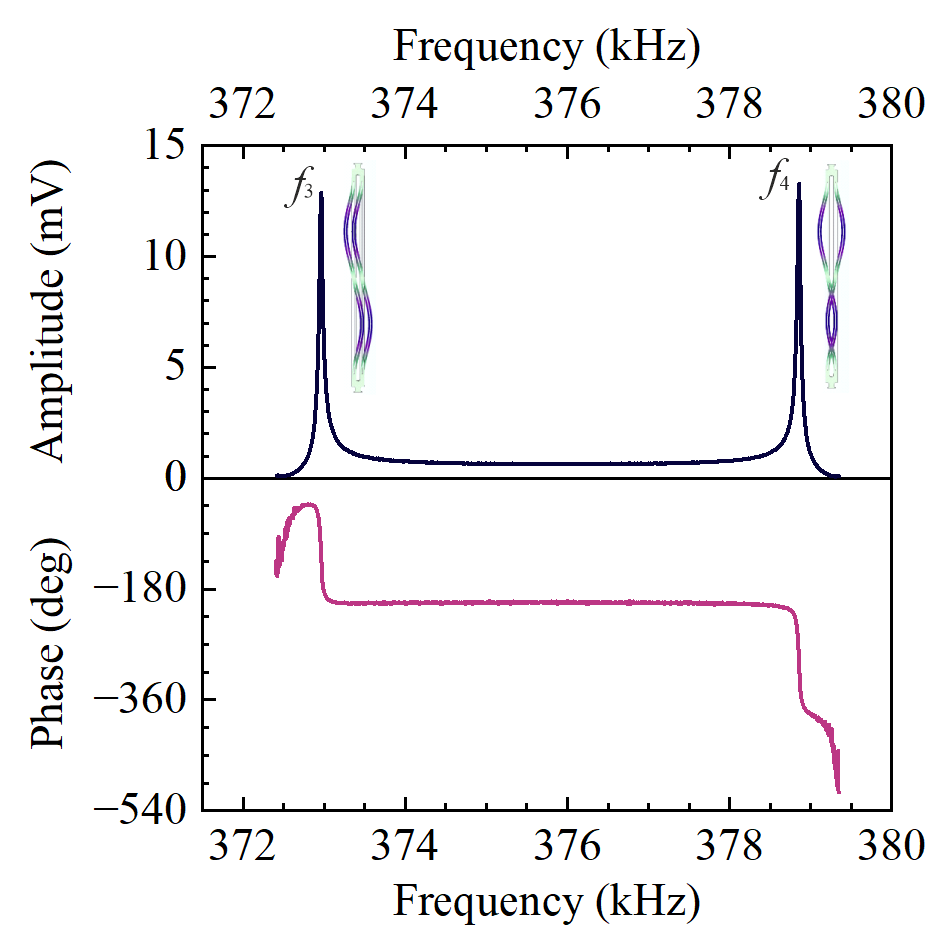}
    \caption{Frequency sweeps for modes $f_{3}$ and $f_{4}$.}
\label{linear_sweeps_f3_f4}
\end{subfigure}
% \vspace{0.1 cm}
\caption{Frequency sweeps showing the modes of interest ($f_{1-4}$) when a drive signal  of amplitude $v_{\text{drive}} = 20$ mV is used, alongside finite-element-analysis simulations of their respective mode shapes. It can be observed that there is negligible feedthrough on mode $f_{1}$ (which is the key mode of interest), and minimal Duffing nonlinearity with the drive amplitude.}
\label{linear_sweeps}
\end{figure}

\begin{figure*}[h!]
    \centering
    \includegraphics[width=0.9\textwidth]{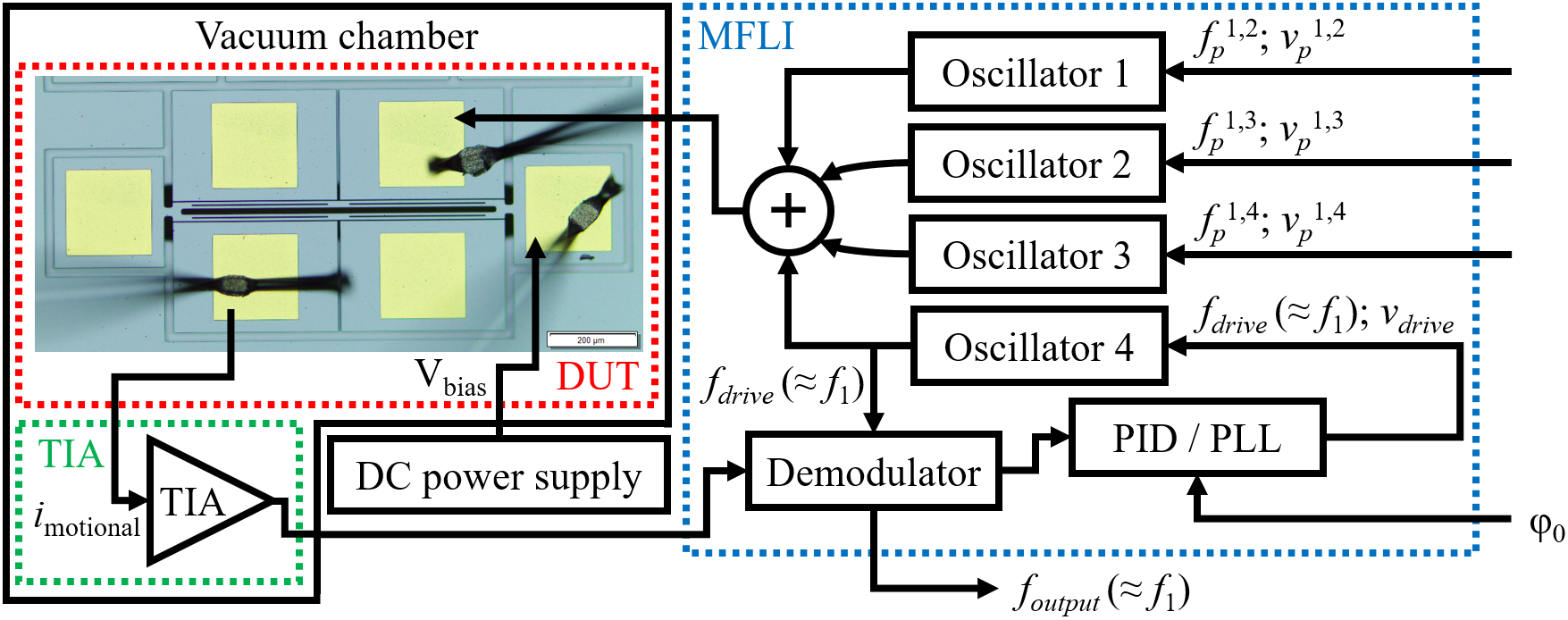}
    \caption{A block diagram of the experimental setup used for characterization of the DUT, with an optical image of the DUT included. The DC power supply provides the bias voltage (kept constant at 30 V throughout) and the TIA converts and amplifies the motional current (with a gain of $\sim$$6$ M$\Omega$). The closed-loop (PID/PLL) mechanism is shown, with the phase set-point ($\varphi_{0}$) as an input, and is used to track the mode of interest ($f_{1}$) which is driven at $v_{\text{drive}} = 20$ mV throughout (oscillator 4). The PMSs are generated by oscillators 1-3, which have controllable frequency and amplitude (i.e., $\Delta f_{p}^{1, j}$ and $\Delta v_{p}^{1, j}$ for $j=2,3,$ and $4$, respectively).}
    \label{experimental_setup}
\end{figure*}

The device-under-test (DUT) is a generic (slotted) double-ended tuning-fork (DETF) MEMS resonator of beam length \SI{780}{\micro\metre}, width \SI{15}{\micro\metre} and thickness \SI{25}{\micro\metre}, which was fabricated using the SOIMUMPS process \cite{duwel2006engineering, zega2018analysis} (as shown in Fig. \ref{experimental_setup}). The DUT is actuated electrostatically and sensed capacitively \cite{algamili2021review}. The beams are slotted to reduce the thermoelastic damping of the resonator \cite{lifshitz2000thermoelastic}. In order to reduce air damping the device is kept under low-vacuum conditions ($\sim$$10^{-1}$ mbar) throughout testing and characterization using a custom-built vacuum chamber. The resonant modes of interest are the flexural modes $f_{1}\approx103.352$ kHz, $f_{2}\approx105.093$ kHz, $f_{3}\approx373.001$ kHz and $f_{4}\approx378.898$ kHz, with Q-factor of $\sim$1.6k, $\sim$1.8k, $\sim$10k and $\sim$11k, respectively (as shown in Fig. \ref{linear_sweeps}). Despite the vacuum conditions, the Q-factors are believed to be limited by air damping, and can be further improved by reducing the vacuum levels to below $10^{-2}$ mbar. The vacuum chamber cavity temperature is kept constant at $20(\pm 0.01)^{\circ}$C using a Thorlabs TEC4015 temperature controller to minimize unwanted temperature-dependent stiffness perturbation \cite{ng2014temperature}.

\subsection{System Setup}
The measurement setup (as shown in Fig. \ref{experimental_setup}) consists of a DC power supply which generates the bias voltage (V\textsubscript{bias} = 30 V), a transimpedance amplifier (TIA) (green-dashed box) which converts and amplifies the output motional current ($i_{\text{motional}}$) into a voltage signal (with a gain of $\sim$$6$ M$\Omega$), and a Zurich Instruments MFLI lock-in amplifier (blue-dashed box) which is used for open-loop and closed-loop characterizations. In addition, the linear actuation signal is generated by Oscillator 4 (see Fig. \ref{experimental_setup}) ($f\textsubscript{drive}\approx f\textsubscript{1}$, with $v_{\text{drive}} = 20$ mV, which is below the critical linear drive amplitude, as no obvious Duffing nonlinearity is observed). Under this testing setup, the maximum amplitudes in the hysteresis experiments (see Fig. \ref{1PM_case_minus100Hz_output_amplitude}, Fig. \ref{vPM1_equals_vPM2_3V_3V_deltafPM1_plus100Hz_amplitude}, and Fig. \ref{3PM_case_amp_response}) did not exceed $\sim$20 mV. These amplitudes are below the maximum linear amplitude of the resonator (without any parametric modulation), of ~25 mV (at which point no obvious Duffing nonlinearity was present for $f_{1}$), as shown in Fig. 2. Therefore, it is believed that the effect of Duffing nonlinearity is at a minimum in this work.

The impact of capacitive feedthrough has also been considered in the experimental setup. By selecting actuation and sense electrodes which are located on either side of the resonator (with the resonator being biased with a DC voltage, V\textsubscript{bias}), so the direct AC current path between the two electrodes is minimized. In addition, we have ensured grounding connections both on the top structural and substrate layers of the device, again minimizing a feedthrough current path between the drive and sense electrodes. Finally, we have chosen a large bias voltage (30 V) to increase the ratio between the motional current signal and the feedthrough. In the experiments, the feedthrough is minimal, especially near the mode of interest, $\sim f_{1}$, as evidenced in the amplitude and phase responses shown in Figs. \ref{linear_sweeps_f1_f2}, \ref{fig:1pmfrequencyresponse}, \ref{fig:2pmfrequencyresponse} and \ref{fig:3pmfrequencyresponse}.

By fully utilizing the "Multi-demodulator Option" on the MFLI, we also generate three additional PMSs ($f_{p}^{1, 2}$ - Oscillator 1, $f_{p}^{1, 3}$ - Oscillator 2, and $f_{p}^{1, 4}$ - Oscillator 3) needed in this experiments. The PMSs used have frequencies determined as follows, and $\Delta f_{p}^{1, j}$ ($j=2,3,$ and $4$) are the control parameters in the experiments:
\begin{align}
\begin{split}
f_{p}^{1, 2}&= f_{2}-f_{1}+\Delta f_{p}^{1, 2}\\
f_{p}^{1, 3}&= f_{3}-f_{1}+\Delta f_{p}^{1, 3}\\
f_{p}^{1, 4}&= f_{4}-f_{1}+\Delta f_{p}^{1, 4}
\end{split}
\end{align}

The amplitudes of the PMSs ($v_{p}^{1, j}$ ($j=2,3,$ and $4$)) are also parameters in the experiments, which can also be controlled via the GUI of the MFLI. For the closed-loop configuration, the PLL and proportional–integral–derivative (PID) controller modules on the MFLI are used to track the frequencies corresponding to the phase set-point ($\varphi_{0}$=$90^\circ$). The MFLI is able to determine the phase difference between the drive and sense signals via a demodulator embedded within, and the PID function within the MFLI automatically corrects the frequency so that the phase difference between the drive and sense signals equal to the $\varphi_{0}$, hence locking onto the frequency corresponding to $\varphi_{0}$. This function of MFLI has been widely used and reported in various MEMS resonant sensors in the community \cite{zhao2019toward, mustafazade2020vibrating}. 

The frequency response behavior in the vicinity of $f_{1}$, i.e., to change the amplitude ratios between the modes as well as the frequency split between the modes \cite{zhao2019toward}, can be altered by adding small detuning terms (i.e., $\Delta f_{p}^{1, j}$) or by changing the amplitude of the PMSs (i.e., $v_{p}^{1, j}$) (for $j= 2, 3, 4$). The PLL/PID phase set-point is arbitrarily chosen to be the phase associated with the peak resonance at $f_{1}$ in the linear case ($\varphi_{0}$ = $90^\circ$). Changing these PMS parameters provides the possibility to tune the hysteresis behavior that is discussed in the experimental results section, however, this is not the focus of this paper, and will be covered in future work. 

\begin{figure}[h!]
\centering
\begin{subfigure}{0.5\textwidth}
    \centering
    \includegraphics[width=0.8\textwidth]{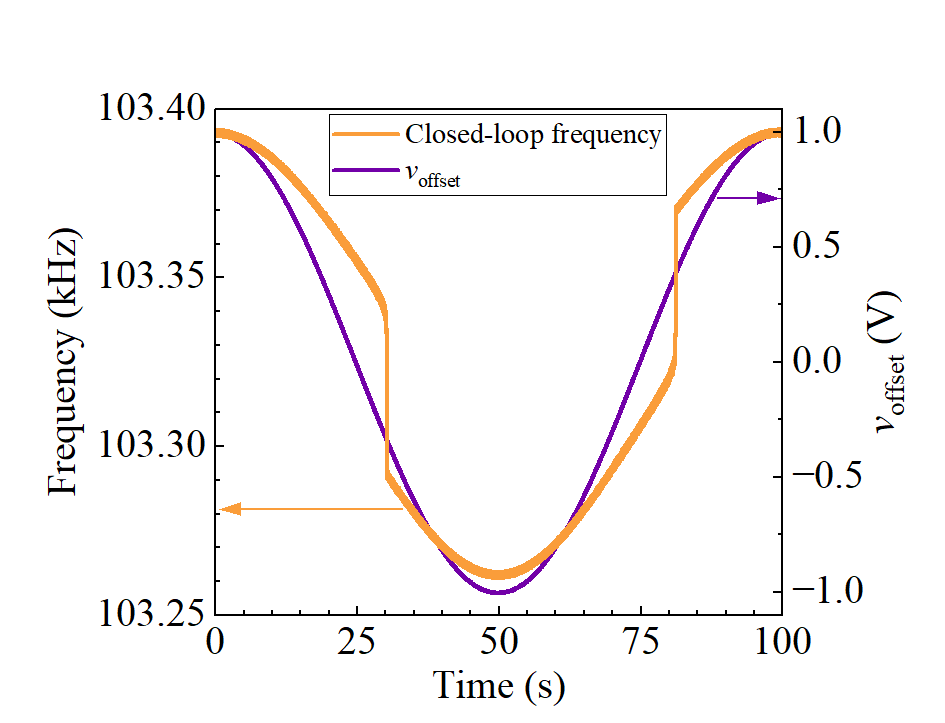}
    \caption{Sinusoidal/quasi-static $v_{\text{offset}}$ sweep}
\label{voffset_and_freq_response_in_time}
\end{subfigure}

\vspace{0.2cm}

\begin{subfigure}{0.5\textwidth}
    \centering
    \includegraphics[width=0.8\textwidth]{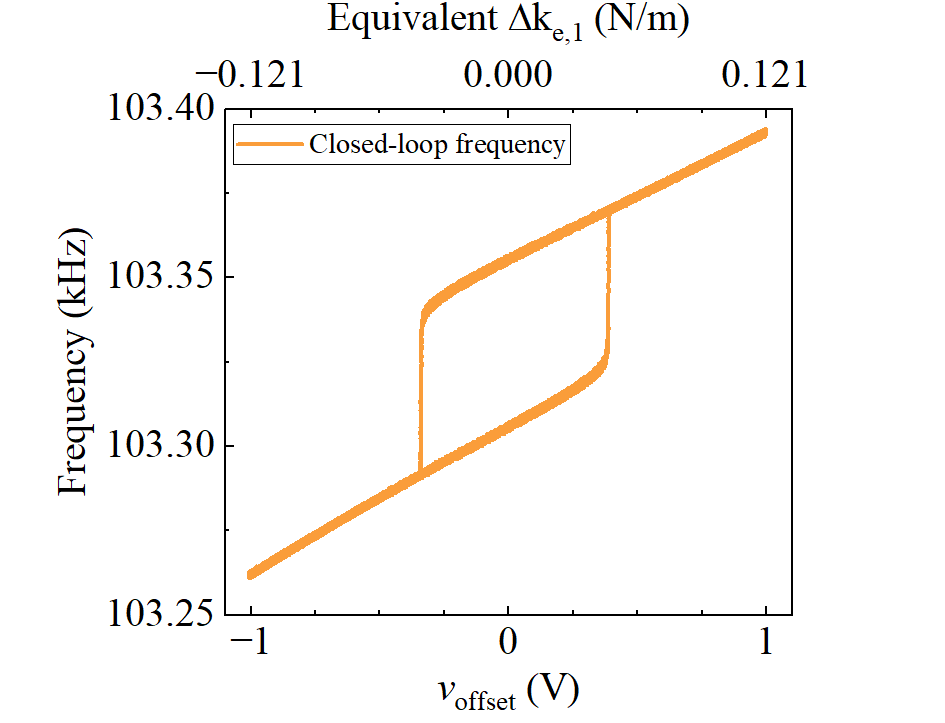}
    \caption{Stiffness ($v_{\text{offset}}$) - frequency space}
\label{freq_vs_offset}
\end{subfigure}
% \vspace{0.1 cm}
\caption{The (a) sinusoidal/quasi-static (10 mHz) $v_{\text{offset}}$ sweep (purple) with the frequency response of the DUT about $f_{1}$ (orange) plotted against time, and (b) the relationship between the frequency response and $v_{\text{offset}}$ (or theoretical equivalent stiffness perturbation $\Delta k_{e, 1}$).}
\end{figure}

For the hysteresis loop characterizations, we introduced a slowly-varying (10 mHz) AC offset voltage ($v_{\text{offset}}$) (added to the MFLI output), which is well below the PLL bandwidth of 20 Hz, as perturbation. The same sinusoidal (10 mHz) $v_{\text{offset}}$ signal is used throughout as the stiffness perturbation. This allows sufficient time for the PLL to respond and is considered a good representation of a static/quasi-static response of the system. Both the perturbation signal ($v_{\text{offset}}$) and the demodulated frequency (and amplitude) responses of the DUT are recorded simultaneously (see Fig. \ref{voffset_and_freq_response_in_time}). The frequency (amplitude) responses are plotted against $\Delta v_{\text{offset}}$ (or the theoretical equivalent $\Delta k_{e, 1}$), with one example shown in Fig. \ref{freq_vs_offset}. 

\section{Experimental Results}
\subsection{Single Parametric Modulation Signal}
\subsubsection{Explanation of hysteresis}
\begin{figure}[h!]
    \centering
    \includegraphics[width=0.4\textwidth]{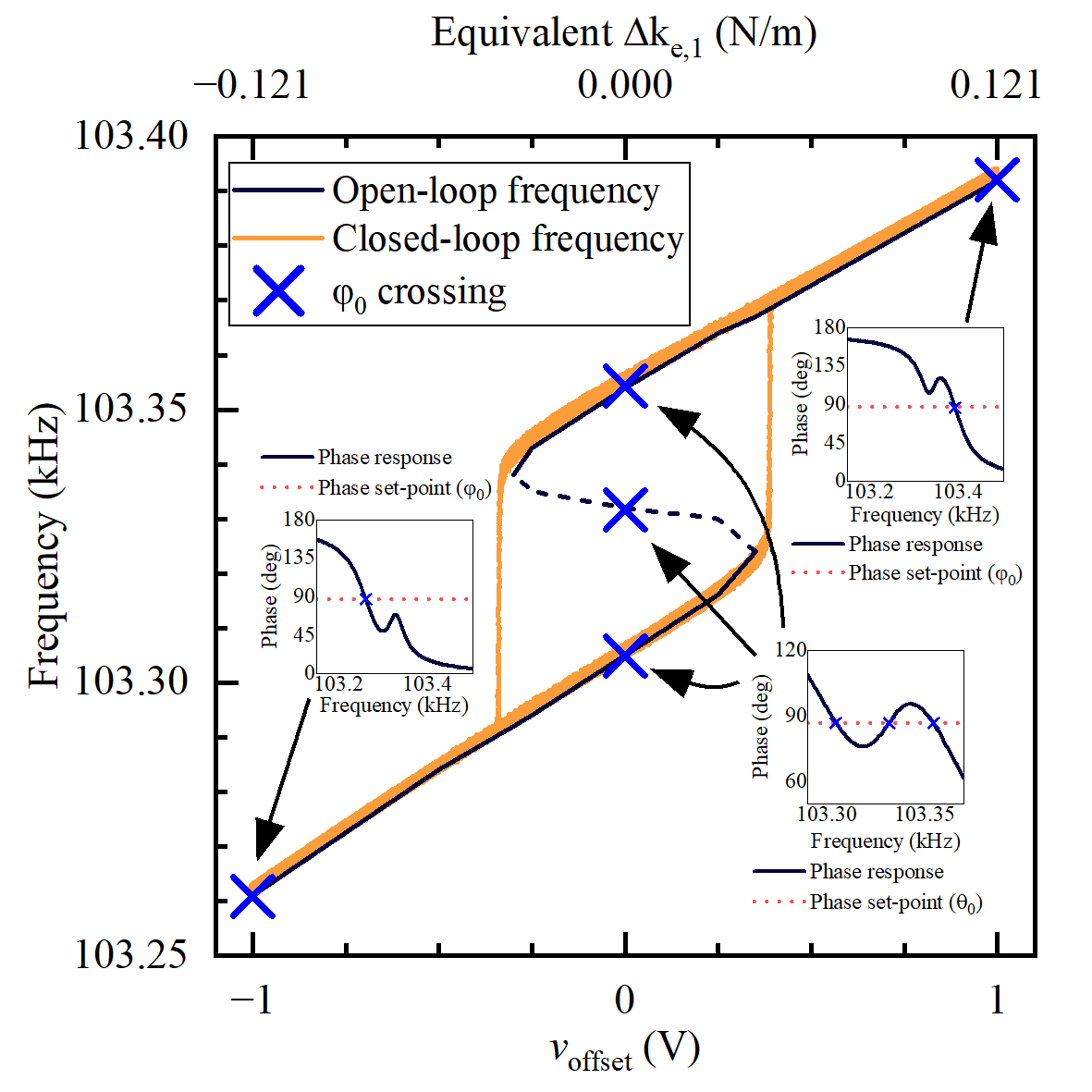}
    \caption{The open-loop (black) and closed-loop (orange) frequency responses when changing the offset voltage (sinusoidal wave, 10 mHz) (or equivalent theoretical stiffness change), where the crosses (blue) show the locations where the open-loop frequency response intersects the phase set-point ($\varphi$\textsubscript{0}). The phase responses corresponding to three offset voltages are inset, which show the phase crossing points.}
    \label{phase_transition_explanation}
\end{figure}

As explained in Section II-A, and reported in previous work, when a PMS ($f_{p}^{1, j}$) is applied, a virtual coupling is created between the first mode and the sideband of the other (higher-order) mode (e.g., $j= 2, 3, 4$). This creates a typical two-mode coupling behavior near the mode of interest ($f_{1}$). With respect to phase, this creates an additional bump/undulation in the phase transition from $180^{\circ}$ to $0^{\circ}$ compared to the conventional phase transition observed in linear systems (see insets of Fig. \ref{phase_transition_explanation}).

As a stiffness perturbation is introduced, the open-loop phase response around the mode of interest ($f_{1}$) shifts (see insets of Fig. \ref{phase_transition_explanation}) - obtained by sweeping the frequency around $f_{1}$. More specifically, it is observed that the number of times the phase response (black lines in insets) intersect the phase set-point, $\varphi_{0}$ (red dotted lines in insets) changes as we sweep across a range of stiffness ($v_{\text{offset}}$) values. The locations of these phase crossing points are indicated (blue crosses in insets) for $v_{\text{offset}}$ values of -1 V, 0 V and 1 V to demonstrate this, showing the corresponding frequencies of these phase crossings. These phase crossing/intersection points are mapped onto the open-loop response (black/black-dashed lines) (see Fig. \ref{phase_transition_explanation}). The occurrence of this behavior can be explained by the linear dynamics of a two-mode-coupled resonator \cite{zhou2023higher}. 

The described behavior, shown in Fig. \ref{phase_transition_explanation} (black lines), is similar to the well-known Duffing nonlinearity, in the sense that one input value can be mapped onto one to three output values \cite{elshurafa2011nonlinear}. In Duffing nonlinearity, the input value is the drive frequency, and the output value is the amplitude, whereas in this case, the input value is the stiffness perturbation ($\Delta k_{e, 1}$) or $v_{\text{offset}}$, and the output is the frequency (or the amplitude).

In a closed-loop configuration a PLL is used to track the frequency of the phase crossings, as we sweep the value of stiffness perturbation (via $v_{\text{offset}}$). This sweeping process is again comparable to frequency sweeps of resonators with Duffing nonlinearity, which also create hysteresis loops. In a similar manner to Duffing nonlinearity, discontinuities will also occur here when the degeneracy of phase crossing points occurs, i.e., when the number of phase crossing points reduces from three to one in two instances (around $0.3$ V and $-0.3$ V) (see Fig. \ref{phase_transition_explanation}). It is worth pointing out that, when the discontinuity occurs, the PLL automatically switches to the other coupled mode through the virtual coupling generated by the parametric modulation operating scheme employed (e.g., from red to the nearby blue arrow, or vice versa, near $f_{1}$ in Fig. \ref{PM_explanation}), therefore a discontinuity in frequency occurs. 

\begin{figure}[h!]
    \centering
    \includegraphics[width=0.35\textwidth]{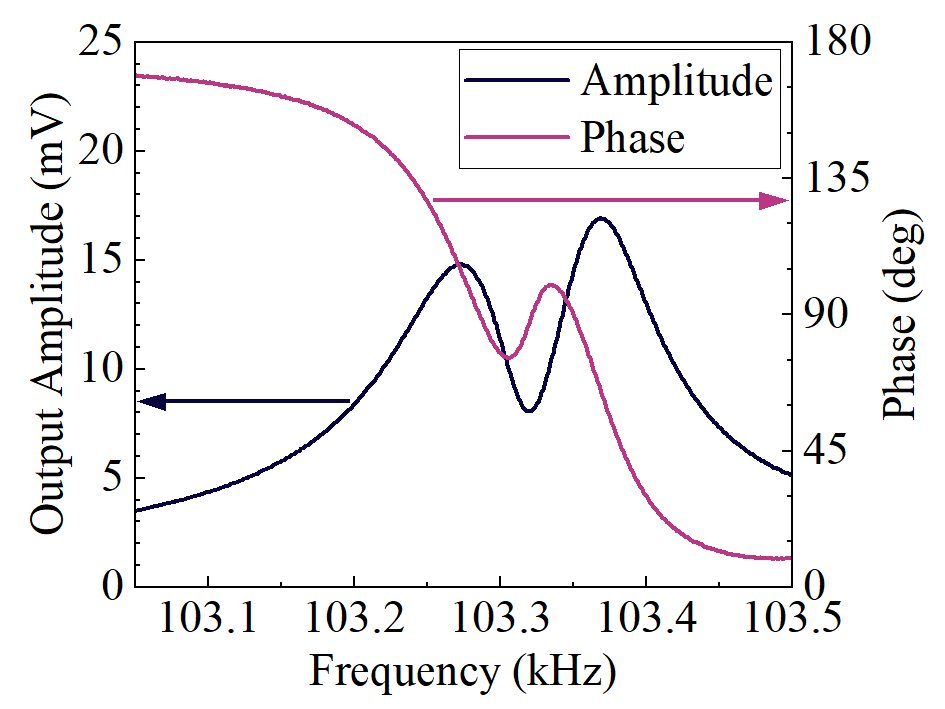}
    \caption{The output amplitude and phase responses when sweeping around $f_{1}$ using one PMS, showing two resonance peaks in the region. The PMS operating parameters used were: $\Delta f_{p}^{1,3} = 0$ Hz and $v_{p}^{1,3} = 3$V.}
    \label{fig:1pmfrequencyresponse}
\end{figure}

\begin{figure}[h!]
    \centering
    \begin{subfigure}{0.45\textwidth}
        \centering
        \hspace*{-1cm}\includegraphics[width=0.8\textwidth]{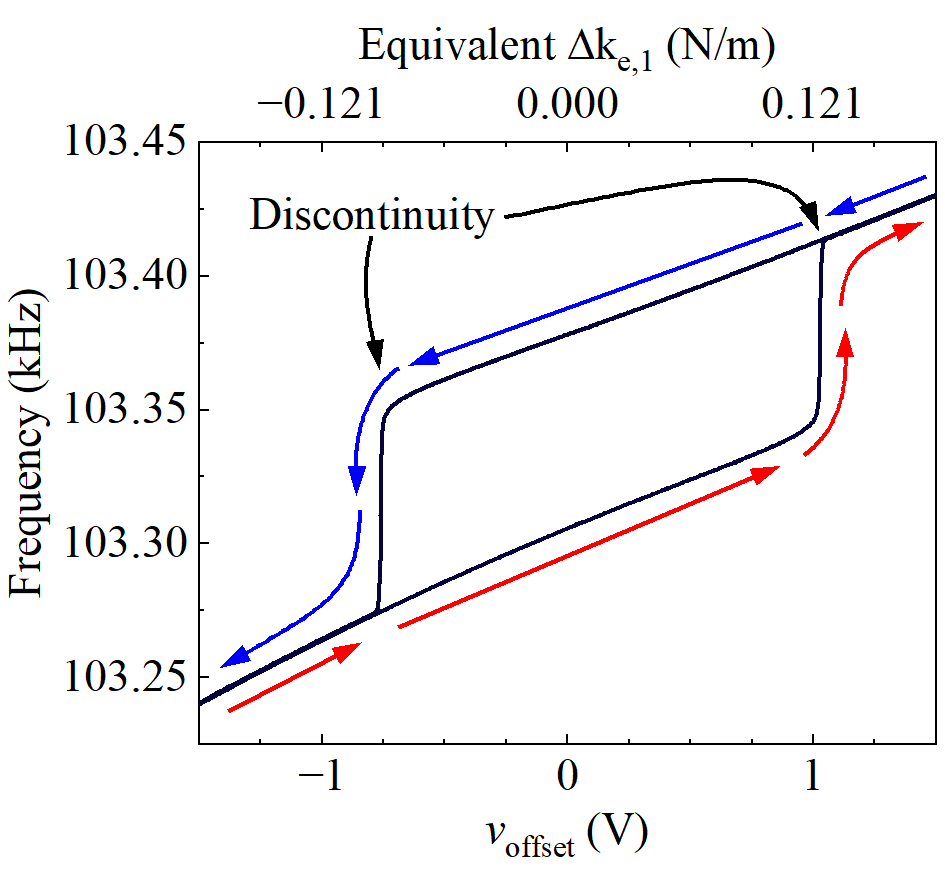}
        \caption{Stiffness perturbation - frequency space}
        \label{1PM_case_minus100Hz_frequency}
    \end{subfigure}%
    \vspace{0.1 cm}
    \begin{subfigure}{0.45\textwidth}
        \centering
        \includegraphics[width=0.8\textwidth]{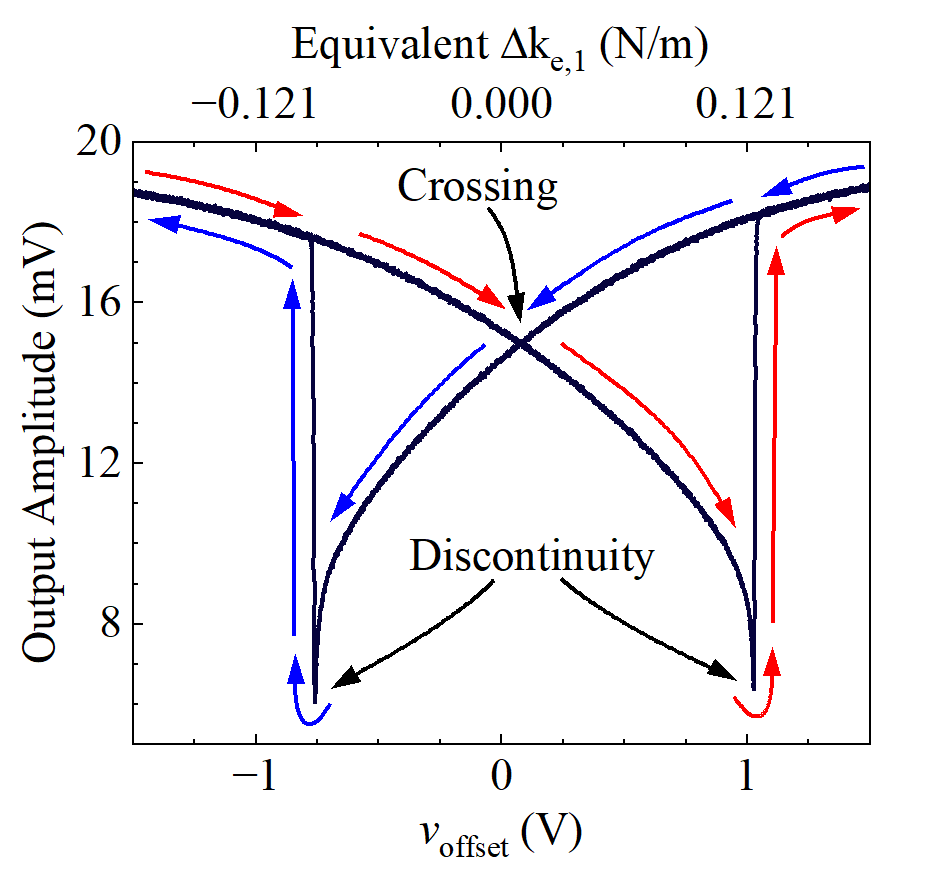}
        \caption{Stiffness perturbation - amplitude space}
        \label{1PM_case_minus100Hz_output_amplitude}
    \end{subfigure}
    \caption{The (a) frequency response - single hysteresis loop, and (b) amplitude response - \emph{pinched hysteresis loop}, when sweeping the offset (sinusoidal wave, 10 mHz) for $v_{p}^{1, 3} = 4$ V and $\Delta f_{p}^{1, 3} = 0$ Hz.}
    \end{figure}

\subsubsection{Two-mode coupling behavior}
An example open-loop frequency sweep showing a representative two-mode coupling behavior, with 1 PMS applied, is shown in Fig. \ref{fig:1pmfrequencyresponse}. The mode-splitting generates two separate peaks in the amplitude response, and a more complex phase transition compared to the linear case (i.e., no PMS employed). This is similar to the two-mode coupling behavior previously reported in \cite{li2021enhancing}.

\subsubsection{Hysteresis loop in stiffness perturbation-frequency space}
The measured hysteresis loop in the frequency response when $v_{p}^{1, 3} = 4$ V and $\Delta f_{p}^{1, 3} = 0$ Hz is shown in Fig. \ref{1PM_case_minus100Hz_frequency}. It is observed that as $v_{\text{offset}}$ is increased (i.e., swept upwards), the frequency increases following the red arrows. When $v_{\text{offset}}$ is approximately $1$ V, a sharp upward transition (discontinuity) is observed. Conversely, when $v_{\text{offset}}$ is swept downwards, the frequency response follows the blue arrows where another discontinuity (downward transition) is observed when $v_{\text{offset}}$ is approximately $-0.75$ V. The results agree well with the simulation results shown in Fig. \ref{sim_1PM_freq_resp}.

\subsubsection{Pinched hysteresis in stiffness perturbation-amplitude space}
It should be pointed out that discontinuity also exists in the stiffness perturbation ($v_{\text{offset}}$) - amplitude space. The output amplitude response with $v_{p}^{1, 3} = 4$ V and $\Delta f_{p}^{1, 3} = 0$ Hz is shown in Fig. \ref{1PM_case_minus100Hz_output_amplitude}. This is commensurate with theoretical calculations shown in Fig. \ref{sim_1PM_amp_resp}. The path again follows the red arrows for the upward sweep and the blue arrows for the downward sweep. It is observed that these paths cross when $v_{\text{offset}}$ is approximately $0$ V. Despite crossing, the path of the response does not switch branches and will continue along the same branch until a discontinuity occurs at $v_{\text{offset}}$ is approximately $1$ V or $-0.75$ V (for the upward or downward branches, respectively), as described for the frequency response. As defined in \cite{chua2014if}, this is a \emph{pinched hysteresis} that has branches which cross paths, however, transitions between branches cannot occur at these branch crossing locations. This also shows that the response of the resonator to an external stimulus, e.g., sensitivity to stiffness perturbation (cf. resistance in memristors), can be switched between two states i.e., following the red or blue arrows, depending on whether the past input value has triggered a state shift. This implies that the sensor has a form of memory \cite{chua2019resistance}, albeit a volatile type of memory at this stage.

The change in amplitude before a discontinuity can be explained by the theory of mode localization \cite{zhao2016review}, where the amplitude is a function of changes in stiffness ($v_{\text{offset}}$). The discontinuity/state shift is due to mode switching resulting from the phase crossing degeneracy, as described earlier. Hence, the difference in the upward and downward sweep curves is due to the different amplitude responses of the modes to changes in stiffness ($v_{\text{offset}}$) \cite{zhao2016review}. 

\subsection{Two Parametric Modulation Signals}% 2PM case
\subsubsection{Three-mode coupling behavior}
\begin{figure}[h!]
    \centering
    \includegraphics[width=0.35\textwidth]{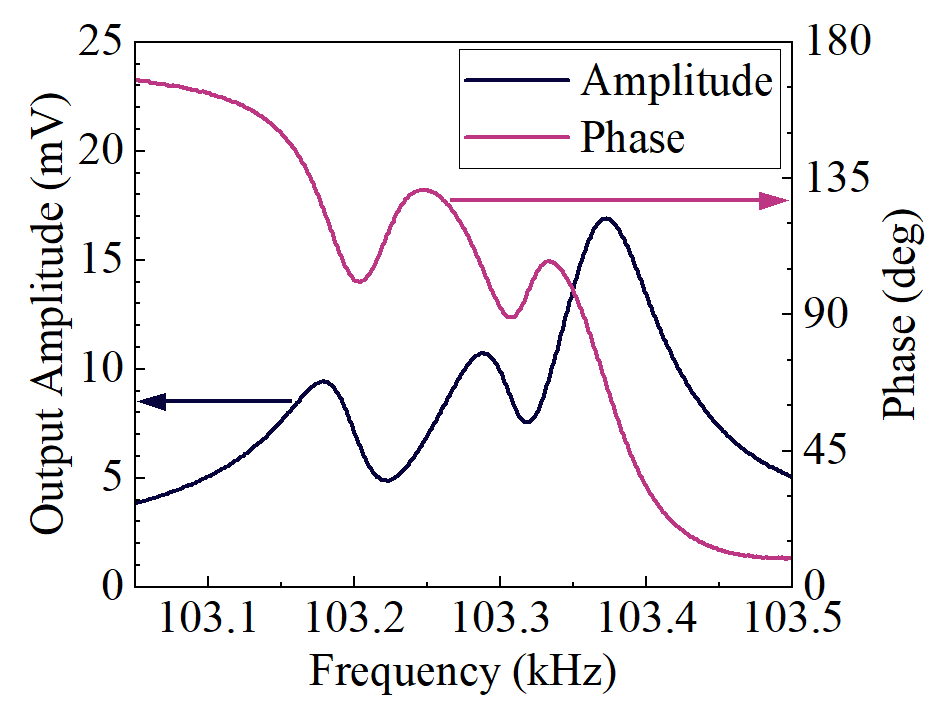}
    \caption{The output amplitude and phase responses when sweeping around $f_{1}$ using two PMSs, showing three resonance peaks in the region. The PMSs operating parameters used were: $\Delta f_{p}^{1,2} = 100$ Hz, $\Delta f_{p}^{1,3} = 0$ Hz and $v_{p}^{1,2} = v_{p}^{1,3} = 3$V.}
    \label{fig:2pmfrequencyresponse}
\end{figure}

An example open-loop amplitude and phase response when two PMSs are employed is shown in Fig. \ref{fig:2pmfrequencyresponse}, where mode-splitting is observed in the amplitude response (i.e., three separate peaks) and a more complex phase response in the phase transition from $180^{\circ}$ to $0^{\circ}$ is observed due to the three-mode coupling behavior. As in the single PMS case, the frequency locations and subsequent number of phase crossings will change as the stiffness perturbation ($v_{\text{offset}}$) as a result of the virtual coupling, as discussed in Section IV-A.

\subsubsection{Pinched hysteresis in stiffness perturbation-frequency space}
\begin{figure}[h!]
    \centering
    \begin{subfigure}{0.45\textwidth}
        \centering
        \hspace*{-1cm}\includegraphics[width=0.8\textwidth]{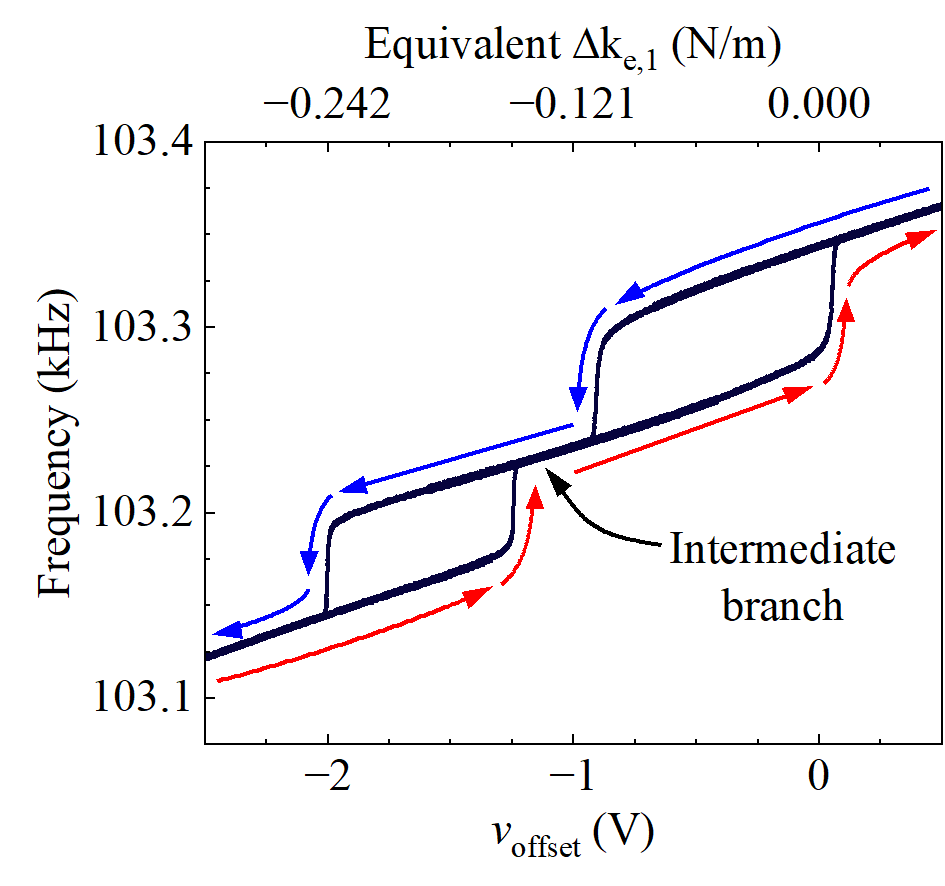}
        \caption{Stiffness perturbation - Frequency space}
        \label{vPM1_equals_vPM2_3V_3V_deltafPM1_plus100Hz_frequency}
    \end{subfigure}%
\vspace{0.1 cm}    
    \begin{subfigure}{0.45\textwidth}
        \centering
        \includegraphics[width=0.8\textwidth]{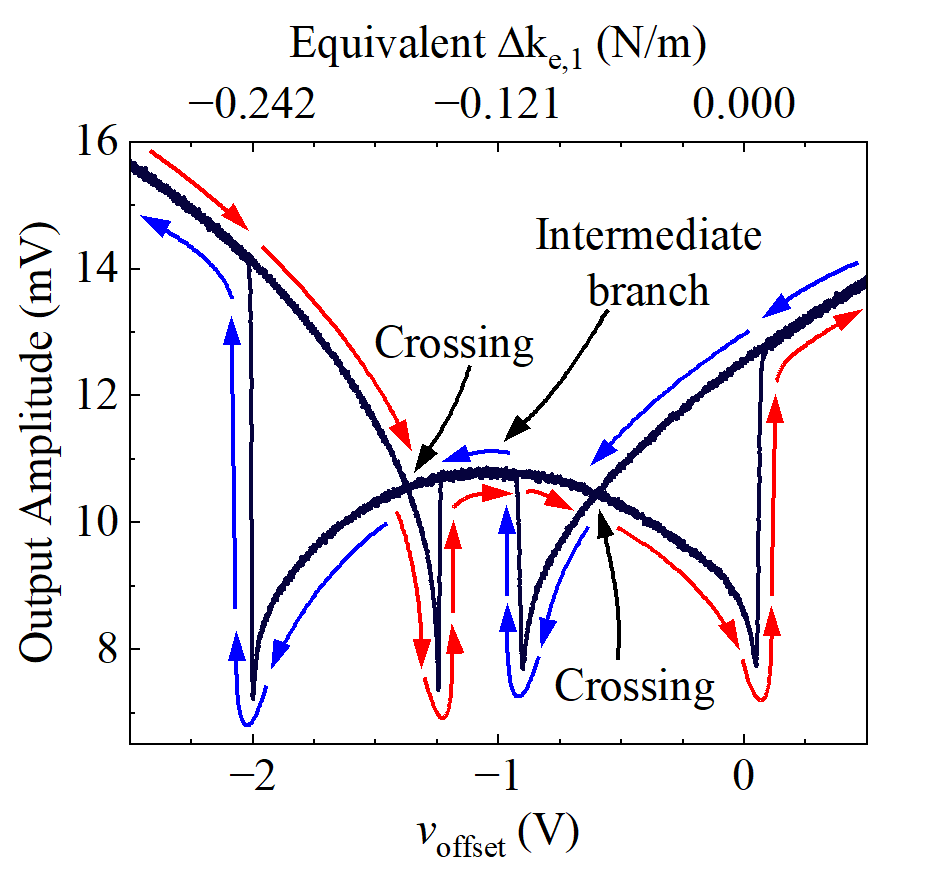}
        \caption{Stiffness perturbation - Amplitude space}
        \label{vPM1_equals_vPM2_3V_3V_deltafPM1_plus100Hz_amplitude}
    \end{subfigure}%
\caption{The (a) frequency response - \emph{pinched hysteresis} and (b) amplitude response - two distinct \emph{pinched hysteresis loops}, when sweeping the offset (sinusoidal wave, 10 mHz) for $v_{p}^{1, 2} = v_{p}^{1, 3} = 3$ V, $\Delta f_{p}^{1, 2} = 0$ Hz and $\Delta f_{p}^{1, 3} = 100$ Hz.}
\end{figure}

Using a closed-loop configuration with two PMSs ($f_{p}^{1, 2}$ and $f_{p}^{1, 3}$) employed, a pinched hysteresis is observed in the frequency response, where the red (blue) arrows indicate the direction of the upward (downward) sweep (as shown in Fig. \ref{vPM1_equals_vPM2_3V_3V_deltafPM1_plus100Hz_frequency}). Here, discontinuities/transitions once again occur for distinct $v_{\text{offset}}$ values depending on the previous PLL frequency. 

For instance, for the upward sweep (red arrows) the PLL frequency increases approximately linearly along the lower branch from a starting point of $v_{\text{offset}}$ = -2.5 V until $v_{\text{offset}}$ = -1.25 V, where an upward discontinuity/transition is observed as the mode that the PLL is tracking switches. Continuing to sweep up, the PLL frequency increases approximately linearly again until $v_{\text{offset}}$ = 0 V, where another upward discontinuity/transition occurs as the mode that the PLL is tracking switches again. Continuing the upward sweep reveals another linear region of the response. Similarly, for the downward sweep (blue arrows) the PLL frequency decreases approximately linearly along the upper branch from $v_{\text{offset}}$ = 0.5 V until $v_{\text{offset}}$ = -1 V, where a downward discontinuity/transition is observed as the mode that the PLL is tracking switches. Continuing to sweep down, the PLL frequency decreases approximately linearly until $v_{\text{offset}}$ = -2 V, where another downward discontinuity/transition occurs as the mode that the PLL is tracking switches again. This brings the response back to the starting point of $v_{\text{offset}}$ = -2.5 V after another linear region of the response. It should be noted that the intermediate branch (see Fig. \ref{vPM1_equals_vPM2_3V_3V_deltafPM1_plus100Hz_frequency}) can be accessed via the upward or downward sweep, and that once on this branch the response is approximately linear. The sweep may continue upward (red arrows) or downwards (blue arrows) from here until reaching a discontinuity/transition at $v_{\text{offset}}$ = 0 V or $v_{\text{offset}}$ = -2 V, respectively. As such, this system can be viewed as a three branch system (the lower, upper and intermediate branches), corresponding to the three modes shown in Fig. \ref{fig:2pmfrequencyresponse}, thus creating two hysteresis loops i.e. one hysteresis loop due to the jumping between the first and the second mode, and the second hysteresis loop due to the jumping between the second and the third mode. When the parameters of the PMSs are chosen in such a way that the two hysteresis loops are separated from each other (i.e., distinct), the phenomenon that satisfies the definition of \emph{pinched hysteresis} \cite{chua2014if} occurs, even though a \emph{pinched} intermediate branch, rather than a crossing, exists. In the experiments, we noticed that this behavior can be tuned by changing the value of $v_{p}^{1, 2}$ and $v_{p}^{1, 3}$. We will leave detailed discussions regarding the tunability of the pinched hysteresis behavior for a follow-up paper.

\subsubsection{Multiple pinched hysteresis loops in stiffness perturbation-amplitude space}
For the output amplitude response MPH are observed (see Fig. \ref{vPM1_equals_vPM2_3V_3V_deltafPM1_plus100Hz_amplitude}), where transitions occur at the same $v_{\text{offset}}$ amplitudes as the frequency response described above. In this case, two distinct branch crossing locations exist at approximately $v_{\text{offset}}$ = -1.5 V and $v_{\text{offset}}$ = -0.5 V, which is evidence of two distinct pinched hysteresis loops within the system. As in the one PMS case, transitions between branches cannot occur at these crossing locations. An intermediate branch also exists, which can be accessed from either the upward (red arrows) or downward (blue arrows) sweep. The system will remain on the intermediate branch until reaching a discontinuity/transition at $v_{\text{offset}}$ = 0 V or $v_{\text{offset}}$ = -2 V, respectively. For instance, for the upward sweep (red arrows) the output amplitude decreases from a starting point of $v_{\text{offset}}$ = -2.5 V until $v_{\text{offset}}$ = -1.25 V, where an upward discontinuity/transition is observed as the mode that the PLL is tracking switches. Continuing to sweep up, the output amplitude decreases until $v_{\text{offset}}$ = 0 V, where another upward discontinuity/transition occurs as the mode that the PLL is tracking switches again. Continuing the upward sweep reveals a region of the response which increases in amplitude. Similarly, for the downward sweep (blue arrows) the output amplitude decreases from $v_{\text{offset}}$ = 0.5 V until $v_{\text{offset}}$ = -1 V, where an upward discontinuity/transition is observed as the mode that the PLL is tracking switches. Continuing to sweep down, the output amplitude decreases until $v_{\text{offset}}$ = -2 V, where another upward discontinuity/transition occurs as the mode that the PLL is tracking switches again. This brings the response back to the starting point of $v_{\text{offset}}$ = -2.5 V after another increasing region of the response. It should be noted that the intermediate branch (see Fig. \ref{vPM1_equals_vPM2_3V_3V_deltafPM1_plus100Hz_amplitude}) can be accessed via the upward or downward sweep, and that once on this branch the response can be swept upward or downward towards to next discontinuity/transition. The distinction from the frequency response (see Fig. \ref{vPM1_equals_vPM2_3V_3V_deltafPM1_plus100Hz_frequency}) is that the crossing points (at $v_{\text{offset}}$ = -1.5 V and $v_{\text{offset}}$ = -0.5 V) form multiple pinched hysteresis, rather than the intermediate branch. As mentioned previously, once in the intermediate branch the response can be swept upward or downward towards to next discontinuity/transition, however, transitions between branches cannot occur at the crossing points. Essentially, transitions between branches occur only at discontinuities, not at crossings.

\subsection{Three Parametric Modulation Signals}
\subsubsection{Four-mode coupling behavior}
\begin{figure}[h!]
    \centering
    \includegraphics[width=0.35\textwidth]{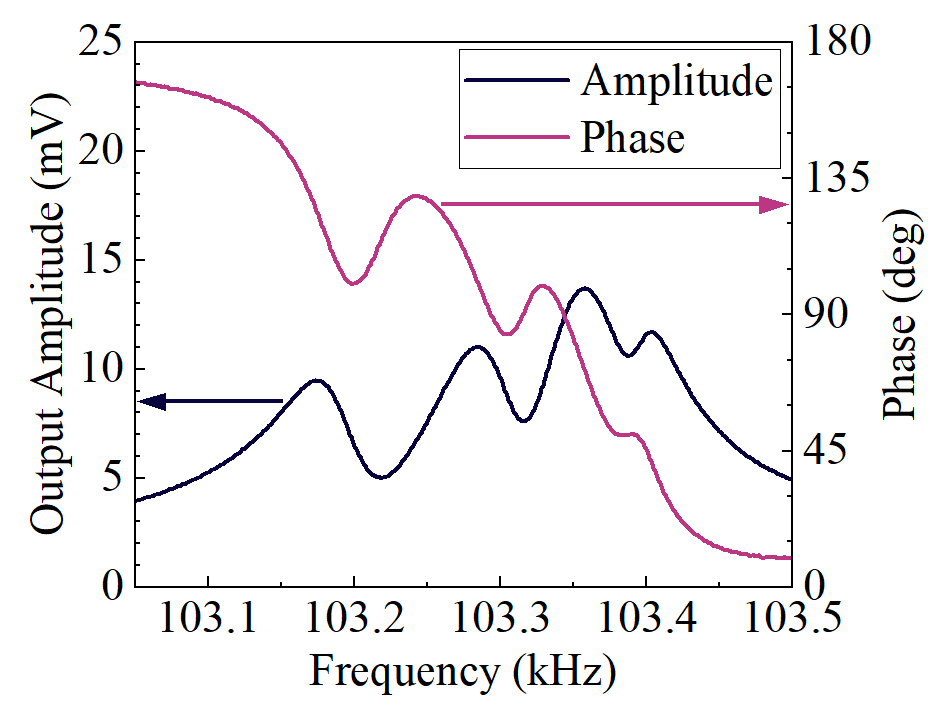}
    \caption{The output amplitude and phase responses when sweeping around $f_{1}$ using three PMSs, showing four resonance peaks in the region. The PMSs operating parameters used were: $\Delta f_{p}^{1,2} = 100$ Hz, $\Delta f_{p}^{1,3} = 0$ Hz, $\Delta f_{p}^{1,4} = -70$ Hz and $v_{p}^{1,2} = v_{p}^{1,3} = v_{p}^{1,4} = 3$V.}
    \label{fig:3pmfrequencyresponse}
\end{figure}

Four-mode coupling behavior is evident from the example open-loop output amplitude and phase response for a three PMS system (see Fig. \ref{fig:3pmfrequencyresponse}), where mode-splitting generates four separate peaks in the amplitude response, and a more complex phase transition from $180^{\circ}$ to $0^{\circ}$.

\subsubsection{Multiple pinched hysteresis loops}

\begin{figure}[h!]
    \centering
    \begin{subfigure}{0.45\textwidth}
        \centering
        \hspace*{-1cm}\includegraphics[width=0.8\textwidth]{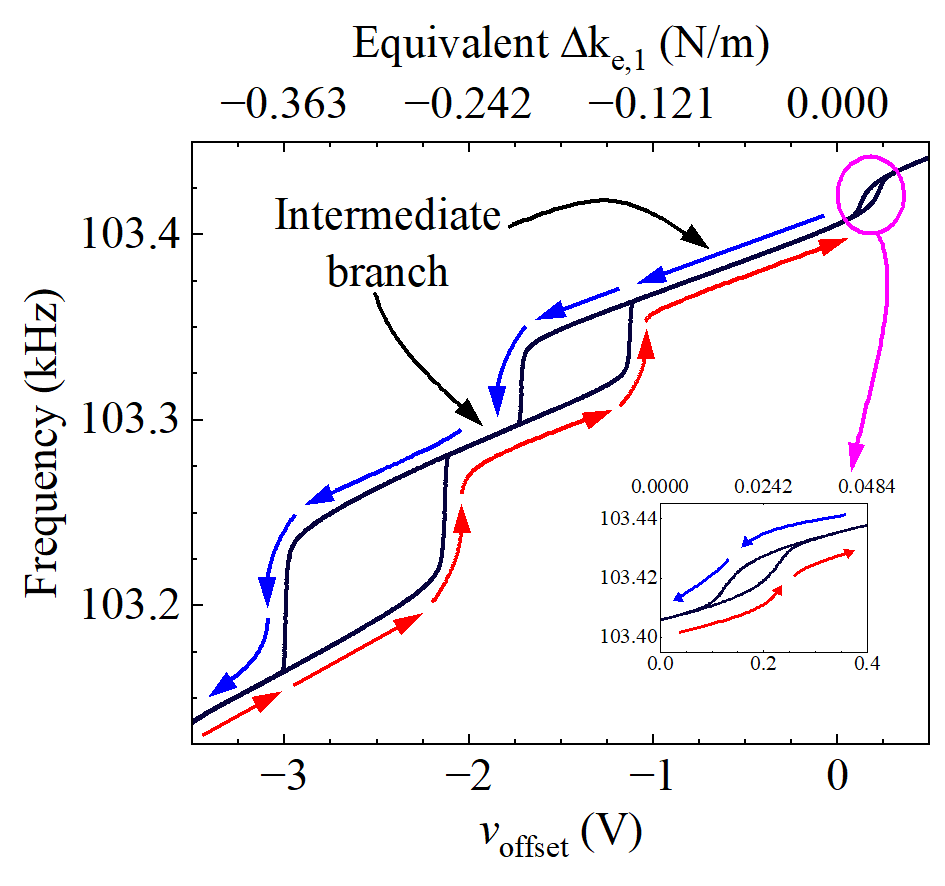}
        \caption{Stiffness perturbation - Frequency space}
        \label{3PM_case_freq_response}
    \end{subfigure}%
\vspace{0.2 cm}    
    \begin{subfigure}{0.45\textwidth}
        \centering
        \includegraphics[width=0.8\textwidth]{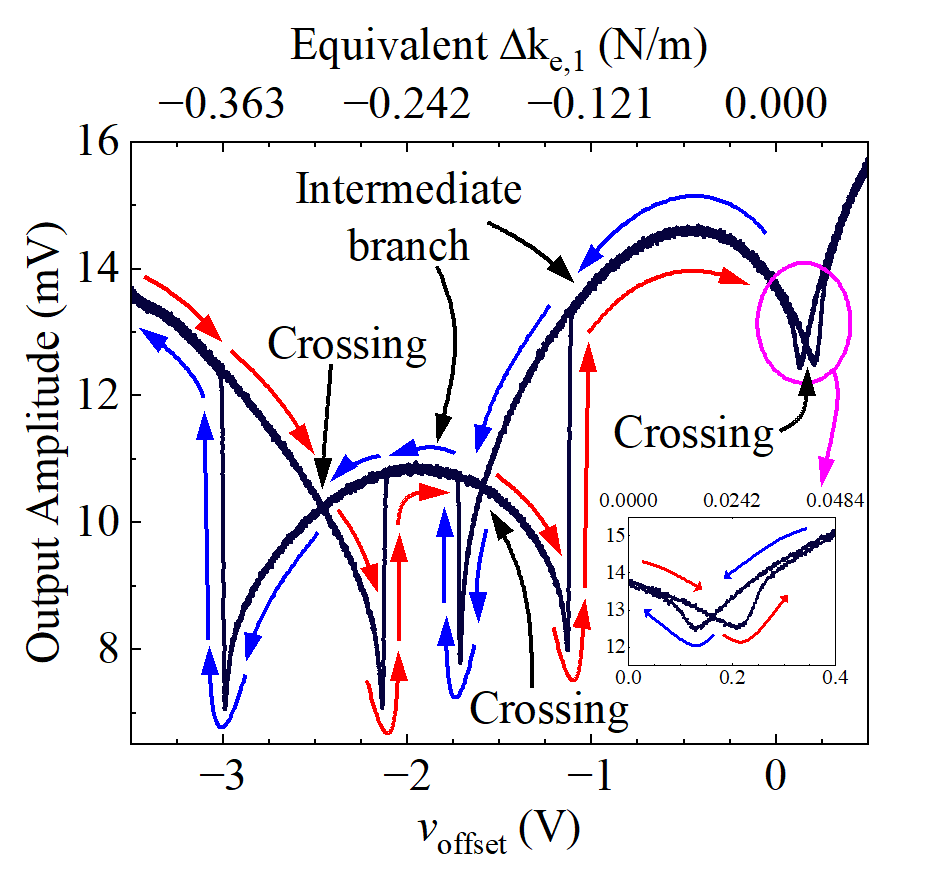}
        \caption{Stiffness perturbation - Amplitude space}
        \label{3PM_case_amp_response}
    \end{subfigure}%
\caption{The (a) frequency response - \emph{pinched hysteresis} with two pinched  intermediate branches, and (b) amplitude response - three distinct \emph{pinched hysteresis} loops with their own pinched crossings, when sweeping the offset (sinusoidal wave, 10 mHz) for $v_{p}^{1, 2} = v_{p}^{1, 3} = v_{p}^{1, 4} = 3$ V, where $\Delta f_{p}^{1, 2} = 97$ Hz, $\Delta f_{p}^{1, 3} = 41$ Hz and $\Delta f_{p}^{1, 4} = -20$ Hz are arbitrarily chosen.}
\end{figure}

Extending this approach, now employing three PMSs ($f_{p}^{1, 2}$, $f_{p}^{1, 3}$ and $f_{p}^{1, 4}$), allows MPH to be observed in both the frequency response and output amplitude response. Following on from the two PMSs case, two intermediate branches are now observed (pinched intermediate branches in the frequency case), which can similarly be accessed and exited in either direction following red (blue) arrows in the upward (downward) direction (as shown in Fig. \ref{3PM_case_freq_response} \& \ref{3PM_case_amp_response}). Discontinuities in the upward sweeps can be observed at $v_{\text{offset}}$ of approximately -2.25 V, -1.25 V and 0.2 V, while in downward sweeps are found at approximately -3 V, -1.75 V and 0.1 V, once again dependent on the previous PLL frequency. It is worth noting that the upward and downward discontinuities in the vicinity of $v_{\text{offset}}=0$ V (see insets of Fig. \ref{3PM_case_freq_response} \& \ref{3PM_case_amp_response}) is comparatively small (0.1 V between the discontinuity points) compared to those observed previously (e.g., one PMS and two PMSs cases) as well as the other discontinuities observed in the three PMSs system discussed here. In the experiments, we noticed that this behavior can be tuned by the PMS parameters, especially the PMS amplitude $v_{p}^{i, j}$ (increasing the PMS amplitude can enlarge the hysteresis loops), which controls the virtual coupling strength between the modes, and hence the frequency difference between the modes \cite{zhou2019dynamic, zhao2019toward}. Even so, the focus of this paper is on presenting the observation of pinched hysteresis, rather than the tunability of the system behavior. This will be investigated in detail in future work.

In the stiffness perturbation-amplitude space, MPH are also observed. Several pinched crossings are observed in the output amplitude response, at approximately -2.5 V, -1.5 V and 0.15 V, solidifying the claim that MPH can be observed in such systems.

\section{Conclusion}
In this work, we have reported for the first time the existence of pinched hysteresis behavior in a MEMS resonator device, demonstrating the viability of creating resonant MEMS sensors which incorporate \emph{cross-domain} (e.g., physical input and electrical output) memristor-like properties, i.e., \textit{MemReSensor}. The dynamics of complex virtual coupling generated by the parametric modulation based operating scheme are described and characterized experimentally. MPH systems have also been observed using the same generic (slotted) DETF type MEMS device as the DUT, between a stiffness perturbation input and an electrical output, suggesting that the pinched hysteresis, and MPH, can exist in practical resonant sensors. 

Similar to the ways memristors with \emph{pinched hysteresis} can be used in unconventional computing and AI \cite{mehonic2020memristors}, we anticipate the applications of the reported device and approach in similar settings. The first key observation of this work is that, employing the reported approach, a MEMS sensor can have \emph{memory} of past input. For example, the sensitivity, i.e., slopes of output amplitude change over input change, can switch from negative to positive or vice versa depending on the past values, as shown in Fig. \ref{1PM_case_minus100Hz_output_amplitude}. Considering the fact that MEMS resonators with nonlinearity and hysteresis can be readily used for reservoir computing, a type of neuromorphic computing that can execute temporal data pattern recognition and classification tasks \cite{caremel2024hysteretic, guo2024mems}, we now expect MEMS resonators to be able to integrate \emph{memory}, sensing, and computation all in one device. This can potentially lead to powerful edge computation devices. The second key observation is that the resonant sensor can have multiple sensitivity values (cf. conductance in memristors), which can be positive, negative, or zero valued depending on the branches (e.g., the slope can be zero on the intermediate branches) and the past values, as shown in Figs. \ref{vPM1_equals_vPM2_3V_3V_deltafPM1_plus100Hz_amplitude} and \ref{3PM_case_amp_response}. This means that we can potentially program the relationship between the physical input and the output. In a similar way to constructing a memristor crossbar array for matrix multiplication \cite{li2023sparse}, we envisage a similar matrix multiplication device is viable by incorporating our approach. The main differences with this MEMS device array are that: (1) the input vector can be composed of multiple physical inputs, rather than voltages, making cross-domain matrix multiplication possible; and (2) the matrix coefficient values, i.e., sensitivity, can be programmed from negative to positive values (including zero), making the matrix more flexible.

Further research will explore the tunability of the pinched hysteresis and MPH presented here, examining whether the locations of discontinuities and pinched crossings, as well as the size of hysteresis loops, can be adjusted for different target application requirements, and whether further memristor-like characteristics, such as spike-timing-dependent plasticity \cite{serrano2013stdp} and nociceptor-like behavior \cite{yoon2018artificial}, can be identified. In addition, we plan to investigate MEMS sensor designs with higher degrees of freedom \cite{zhao2015three}, with the aim of revealing further interesting dynamics and discovering whether multi-modal sensor switching responses can be observed. Future work will also explore the implementation of the observed pinched hysteresis behavior in (multi-)functional sensors to realize a practical cross-domain \emph{MemReSensor}, as well as their application to in-sensor computing and cross-domain matrix multiplication.

\section*{Acknowledgment}
We would like to thank the Henry Royce Institute for access to the Leeds Nanotechnology Cleanroom at the University of Leeds through the Researcher Equipment Access Scheme (EPSRC Grant Number EP/P022464/1). We would also like to thank Ms. Jingqian Xi from Huazhong University of Science and Technology for the useful discussions and insights.

\bibliographystyle{IEEEtran}
\bibliography{IEEEabrv,Bibliography}

\newpage

\section*{Biography}
\vspace{-1cm}
\begin{IEEEbiography}[{\includegraphics[width=1in,height=1.25in,clip,keepaspectratio]{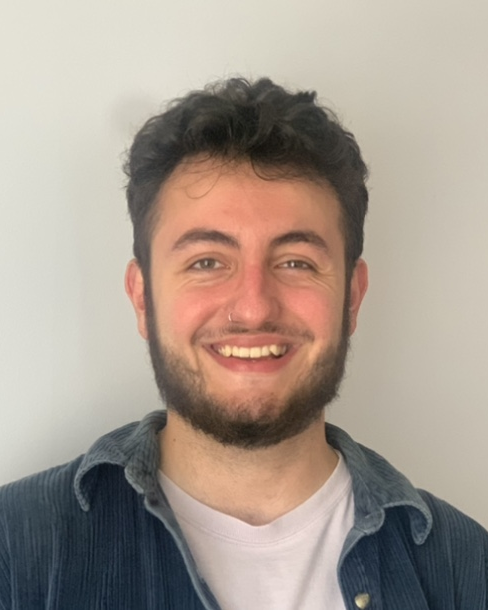}}]{Erion Uka} received the M.Eng. degree in electronic engineering with nanotechnology from the University of York, York, U.K. in 2022. He is currently pursuing the Ph.D. degree in electronic engineering with the School of Physics, Engineering and Technology, University of York, York, U.K. His research interests include resonant MEMS devices, high sensitivity MEMS sensors and unconventional MEMS dynamics, with a focus on new sensing paradigms and application areas.

\vspace{-1cm}

\end{IEEEbiography}
\begin{IEEEbiography}[{\includegraphics[width=1in, height=1.25in,clip,keepaspectratio]{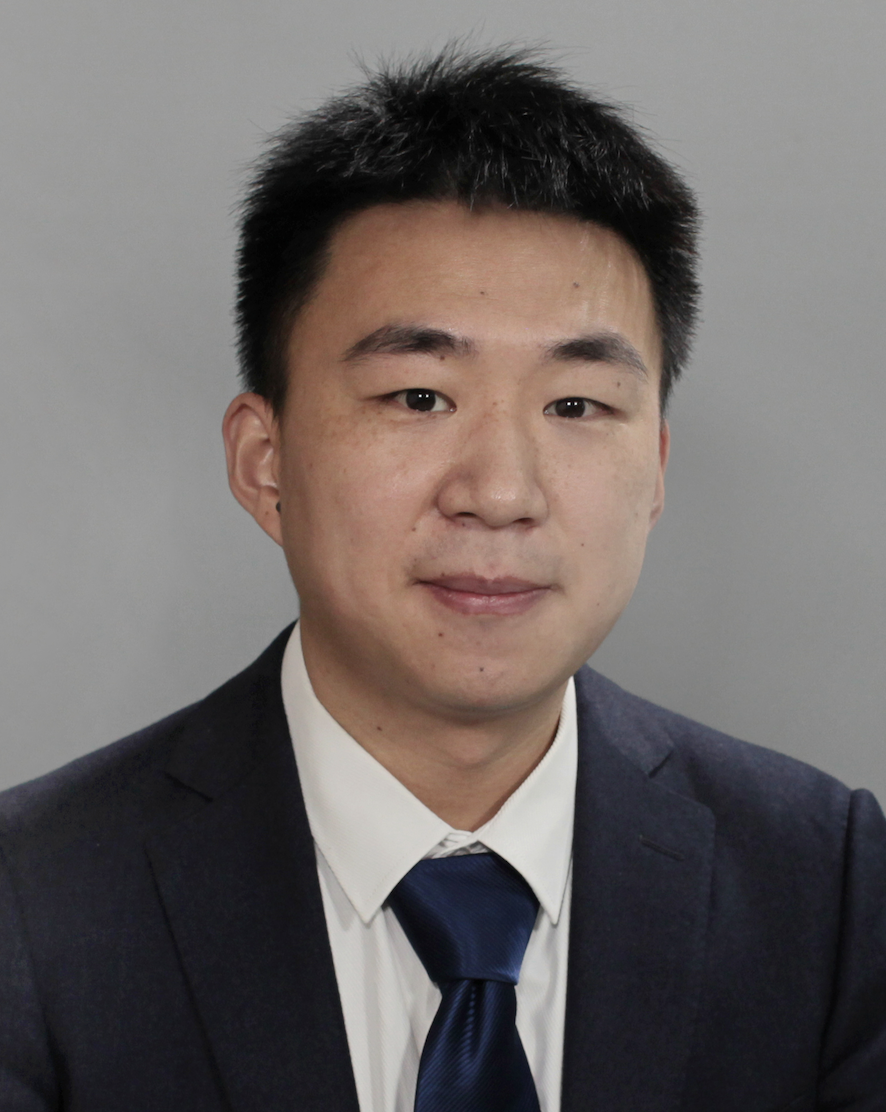}}]{Chun Zhao} (S'14 - M'16 - SM'20) is currently a Lecturer (Assistant Professor) in Microengineering at the University of York, UK. He received B.Eng. degree from the Huazhong University of Science and Technology, Wuhan, China, in 2009; M.Sc. degree from Imperial College London, London, U.K., in 2011; and Ph.D. degree from the University of Southampton, Southampton, U.K., in 2016.

Prior to joining the University of York in 2022, he was a Research Scientist with Sharp Laboratories of Europe, Oxford, UK (2015-2016), a Research Associate in MEMS at the University of Cambridge, UK (2016-2018), and an Associate Professor with the School of Physics, Huazhong University of Science and Technology, Wuhan, China (2018-2021). Dr. Zhao is currently serving as an Associate Editor for IEEE Sensors Journal, and has served in the Technical Program Committee for MEMS 24' and 25'. His research interests include MEMS resonators, high resolution MEMS sensors, unconventional materials and physics for MEMS (e.g. modal interactions and nonlinear effects).
\end{IEEEbiography}

\vfill

\end{document}